\documentclass{PoS}

\title{High sensitivity VLBI with SKA}

\ShortTitle{VLBI with SKA}

\author{\speaker{Cristina Garc\'ia-Mir\'o}, Antonio Chrysostomou \\ Square
Kilometre Array Organisation (SKAO), Jodrell Bank Observatory,\\ Lower Withington, Macclesfield,
Cheshire SK11 9DL, United Kingdom\\ E-mail: \email{c.garcia-miro@skatelescope.org,
a.chrysostomou@skatelescope.org}}

\author{Zsolt Paragi, Ilse van Bemmel\\ Joint Institute for VLBI ERIC (JIVE), \\ Oude
Hoogeveensedijk 4, 7991\,PD Dwingeloo, The Netherlands\\ E-mail: \email{zparagi@jive.eu,
bemmel@jive.eu}}

\abstract{The Square Kilometre Array (SKA), with the aim of achieving a collecting area of one
square kilometre, will be the world's largest radio telescope. A scientific collaboration between 12
countries (with more to join), it will consist of one Observatory with 2 telescopes located in South
Africa and Australia. The telescope deployment is planned in two phases, but even in its first stage
(SKA1) it will already enable transformational science in a broad range of scientific objectives.
The inclusion of SKA1 in the Global VLBI networks (SKA-VLBI) will provide access to very high
angular resolution to SKA science programmes in anticipation of the science to be realized with the
full telescope deployment (SKA2). This contribution provides an overview of the SKA Observatory VLBI
capability, the key operational concepts and outlines the need to update the science use cases.}

\FullConference{14th European VLBI Network Symposium \& Users Meeting (EVN 2018)\\ 8-11 October
2018\\ Granada, Spain}

\begin{document}

\section{The Square Kilometre Array}

The Square Kilometre Array (SKA) will be the world's largest radio telescope ever constructed. It
will be a multi-purpose radio observatory, designed to cover the frequency range from 50 MHz up to
20 GHz, that will play a major role in answering key questions in modern astrophysics and cosmology.
The SKA Observatory will include two very different radio telescopes in terms of the receiving
elements used. The LOW telescope will be located at the Murchison Radio Astronomy Observatory in
Australia and the MID telescope in the Karoo region in South Africa. The headquarters will be
located at the Jodrell Bank Observatory in the United Kingdom.

\begin{figure}[!ht] \vspace{1.0 cm} \includegraphics[angle=0,width=6.2in]{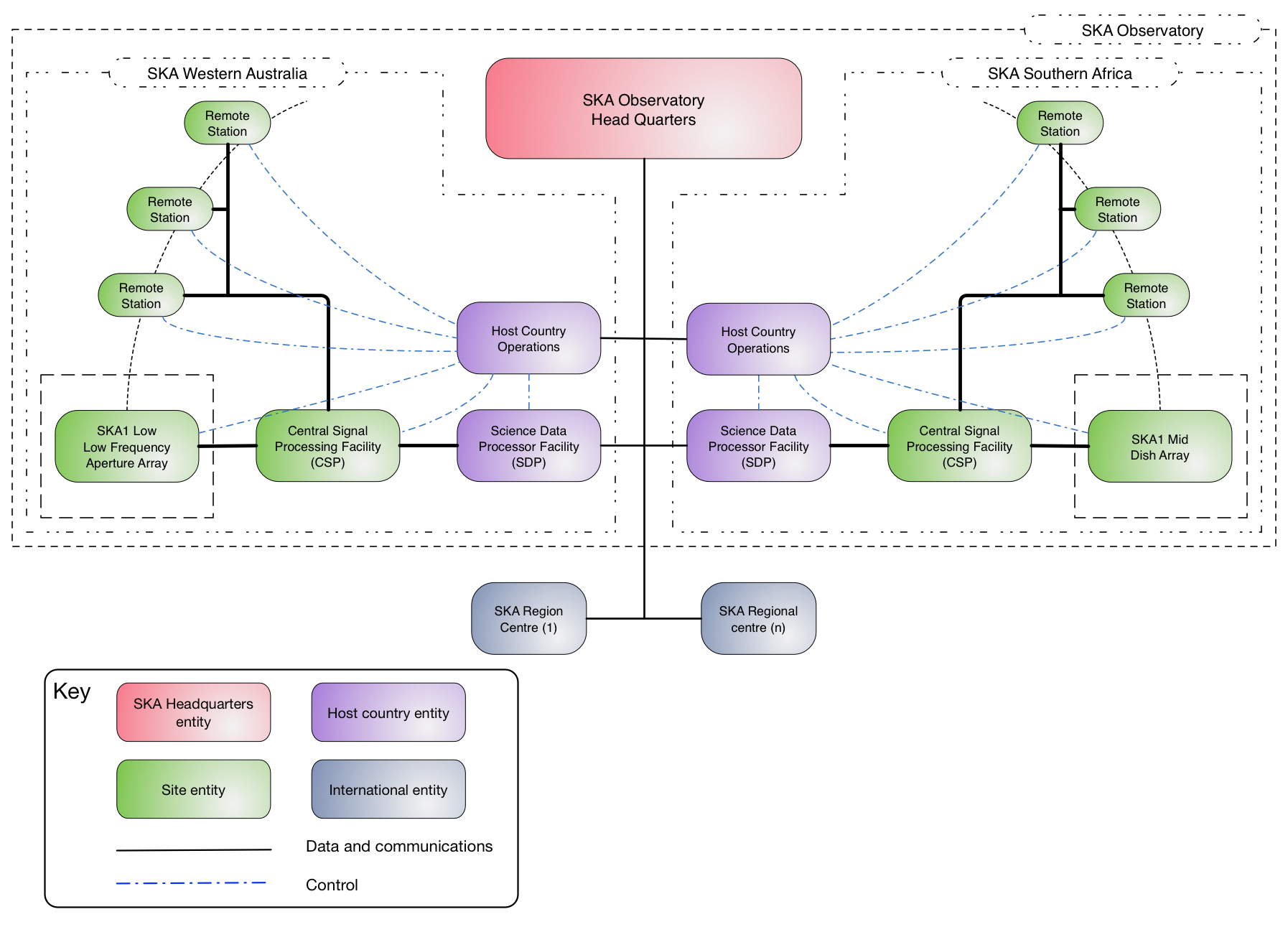} \caption{A
schematic of the SKA Observatory structure.} \label{fig1} \end{figure}

Due to the magnitude of the project it is planned to be deployed in two phases. The design of the
first phase, SKA1, is supported by a broad international collaboration organised into different
Consortia, each responsible for the design of a distinctive component or element of the telescopes,
e.g. dishes, correlator, signal and data transport, etc. (see Fig.~\ref{fig1}). At the time of
writing, this design is being finalised with the Critical Design Reviews (CDRs) of the different
elements, and nearing completion. Once the elements' designs have been evaluated and closed-out, the
Consortia dissolve and a Bridging Phase starts with support from the former participant
institutions. The Bridging Phase prepares for a System Level Critical Design Review to assess the
design of the full SKA1 Observatory, and leads towards construction.

\section{Implementing VLBI at the SKA1}

Comparing to existing radio interferometers, the SKA telescopes will access a wide range of angular
resolutions. In its initial phase, SKA1 will greatly surpass the capability of currently
operational, connected interferometers in terms of sensitivity, survey speed and sub-arcsecond
angular resolution. But in terms of the latter, there are science cases that would benefit from
resolutions that are not achievable with connected-element interferometers. Nowadays the highest
angular resolutions are achieved by Global VLBI observations, which use coordinated networks of
radio telescopes located around the globe and in space, to synthesise an equivalent Earth size
instrument or even larger. Inclusion of SKA1 in the Global VLBI networks (SKA-VLBI) will provide
multi-beam capability with $\mu$Jy sensitivity in N-S baselines with access to the Southern
Hemisphere and simultaneous images of the sky at a broad range of angular resolutions, down to
sub-milliarcseconds for the higher observing frequencies of the SKA (\cite{SKA-VLBI15}, and
references therein). The boost in sensitivity will allow efficient VLBI surveys of the sub-mJy
source population and access to the $\mu$Jy regime for individual sources, with improved fidelity
due to the SKA superior amplitude and polarization calibration. The multi-beam capability will
enable high precision astrometry and enhanced phase referencing techniques, transient localisation,
etc.

The Horizon 2020 JUMPING JIVE project recognised the scientific relevance of the SKA-VLBI. It has a
devoted work package, \lq\lq VLBI with SKA'', to focus on the definition of a SKA-VLBI Operational
Model, define VLBI interfaces and requirements for the SKA, and to help the community develop
Science Use Cases for SKA-VLBI and strategies for possible future SKA-VLBI Key Science Projects
(KSPs) \cite{SKA-VLBI18}. It will provide four main deliverables, scheduled in a period of 42
months. This contribution summarises the outcomes of the first deliverable \lq\lq Details on VLBI
Interfaces to SKA Consortia''. This consists of a description of the VLBI element that will provide
the buffering and streaming capability to send SKA VLBI data to an external VLBI correlator, as well
as all necessary interfaces for scheduling and conducting VLBI observations with the SKA.

%
\section{SKA-VLBI Capability}

VLBI will be an observing mode of the SKA Observatory with the aim of including the participation of
the SKA1 LOW and MID telescopes in Global VLBI observations. In this mode, SKA1 provides multiple
sensitive VLBI tied-array beams for inclusion in VLBI observations. Each SKA1 VLBI beam acts as an
individual element in the VLBI network, equivalent to an individual single beam radio telescope
participating in the observation. Tied-array beams are produced by a beamformer in the SKA
correlator. Each tied-array beam is formed within a subset (or subarray) of LOW stations or MID
dishes by coherently combining the signals from the receptors, such that the combined gain is
directed at a specified point on the sky. At least 4 dual-polarisation VLBI beams are formed from
one or more subarrays, compatible with the standard VLBI observing mode. The SKA Observatory Data
Products for this mode are: VLBI voltage beams formatted into VDIF (VLBI standard for data
interchange format) packets to be correlated at an External Correlator, associated metadata, and
image cubes produced by the Science Data Processor (SDP) from the same subarray from which the VLBI
beams were formed. These simultaneous images will be used to calibrate the VLBI data, as well as to
complement the science data return by providing images of the sky with different bandwidths and
angular and spectral resolutions.

\begin{figure}[!ht] \vspace{1.0 cm} \includegraphics[angle=0,width=6.2in]{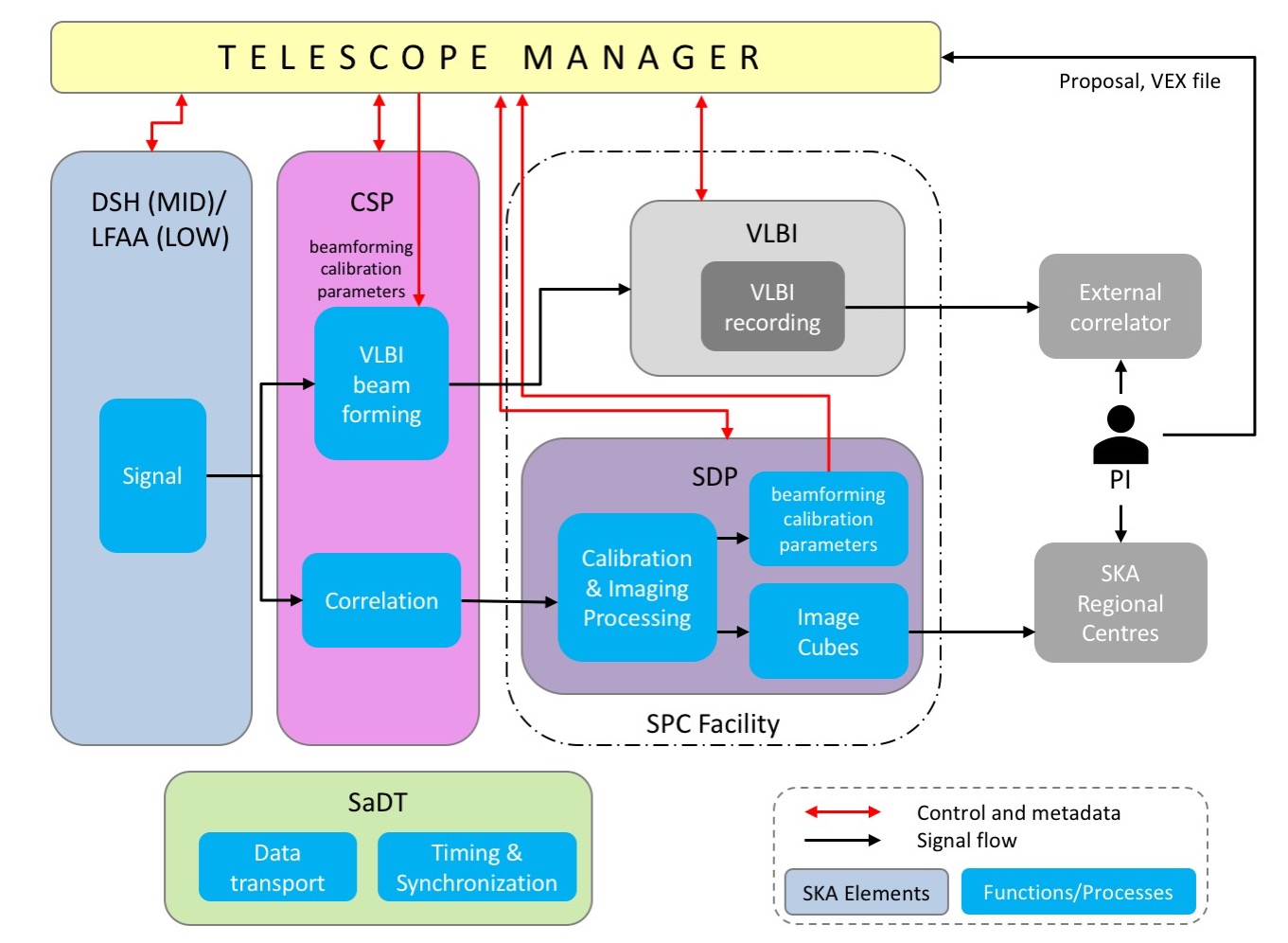}
\caption{VLBI capability integration in the SKA1 Observatory for LOW and MID telescopes.}
\label{fig3} \end{figure}

The key operational concepts of the SKA Observatory \cite{SKA-Ops} that define how the VLBI
operations will be performed are (i) the ability to configure and simultaneously operate independent
subarrays, (ii) the simultaneity of Imaging and VLBI observing modes in the same subarray for
calibration and commensal science and (iii) the independent multi-beam capability for each subarray.
The VLBI Operational workflow will be as follows (Fig.~\ref{fig3}):

\begin{itemize}

\item Approved observing proposals become Observatory Projects. The PI will provide all necessary
information for the observation design, in particular the VEX file (version 2.0) that describes all
the details needed to perform the VLBI observation and to correlate it. With this information the
observation is planned and scheduled, and a Scheduling Block (SB) is generated for observation
execution. These functions are performed by the Telescope Manager (TM) element.

\item Digitised signals from the receivers (LOW log-periodic antennas, or MID dishes) are fed into
the Correlator BeamFormers (CBF) of the Central Signal Processor (CSP), located at each SKA1
telescope site. Apart from complex correlation, the correlators form tied-array beams for VLBI,
Pulsar Timing and Pulsar Search. The VLBI beams are corrected for polarization impurity, channelized
in real representation, and formatted into VDIF packets. Channels that are adversely affected by RFI
are either excised or flagged accordingly.

\item VLBI VDIF packets are sent to the VLBI Terminal located at the Science Processing Centre (SPC)
for either real-time streaming to the External correlator (e-VLBI mode) or for recording (buffering)
and streaming after the observation has completed. The e-VLBI mode has several advantages over the
recorded mode, such as real-time fringe verification and troubleshooting, fast scientific
turnaround, observing configuration adaptability, etc. The VLBI Terminal has access to the
Observatory metadata via subscription to the TM element to generate an observing log in support for
the external correlation and imaging post-processing calibration.

\item Simultaneously, the correlator visibilities are received by the Science Data Processor (SDP)
and processed by the real-time calibration pipeline. This pipeline calculates the beamforming
calibration parameters (delay models, complex gains, polarisation corrections, etc.) that are then
sent to the CBF via the metadata flow managed by TM. The real-time calibration must be determined
and applied to maximise the coherent tied-array beam gain as well as its polarisation purity while
counteracting possible ionospheric position jitter. Correlator visibilities are also processed by
SDP to produce image cubes that are sent to the SKA Regional Centres (SRCs) for further processing
and analysis.

\item In parallel, the Signal and Data Transport (SaDT) element is responsible for the science data
and non-science data communications links, that route VLBI data to the VLBI Terminal from the
correlators and connect with TM for monitor and control tasks. The SaDT element is also responsible
for the realisation of the SKA timescale based on the use of hydrogen masers which provide a
phase-coherent reference signal with the stability required by VLBI.

\item The PI accesses the data products from the external correlator (VLBI visibilities, pipeline
products including VLBI images and metadata, etc.) and from the SRCs (images in support of VLBI
calibration and, if justified in the VLBI proposal, full resolution imaging data products from the
same subarray and/or the whole array).

\end{itemize}

\subsection{VLBI with the SKA1-MID}

The SKA1-MID correlator and beamformer design presents a FX-type FPGA based correlator with a very
flexible architecture that efficiently manages its processing resources \cite{SKA-Base}. It is able
to simultaneously process up to 16 independent subarrays, as well as all observing modes
simultaneously within each subarray. To do so requires a bandwidth sacrifice, especially for Band 5
(4.6-15.3 GHz). The architecture allows input from 200 antennas separated by several 1000s km and
can be easily upgraded to support 20\% more antennas. The input signal from the different observing
bands is divided into 200 MHz frequency chunks, called Frequency Slices, with each slice processed
by a Frequency Slice Processor (FSP) - a set of FPGAs configured to perform correlation or the
different types of tied-array beamforming. The MID correlator is required to form, in different
directions on the sky, 1500 beams for Pulsar Search (PSS), 16 beams for Pulsar Timing (PST) and 4
beams for VLBI. The beams can be produced from the same subarray or distributed amongst different
subarrays.

\begin{figure}[!ht] \vspace{1.0 cm} \includegraphics[angle=0,width=6.2in]{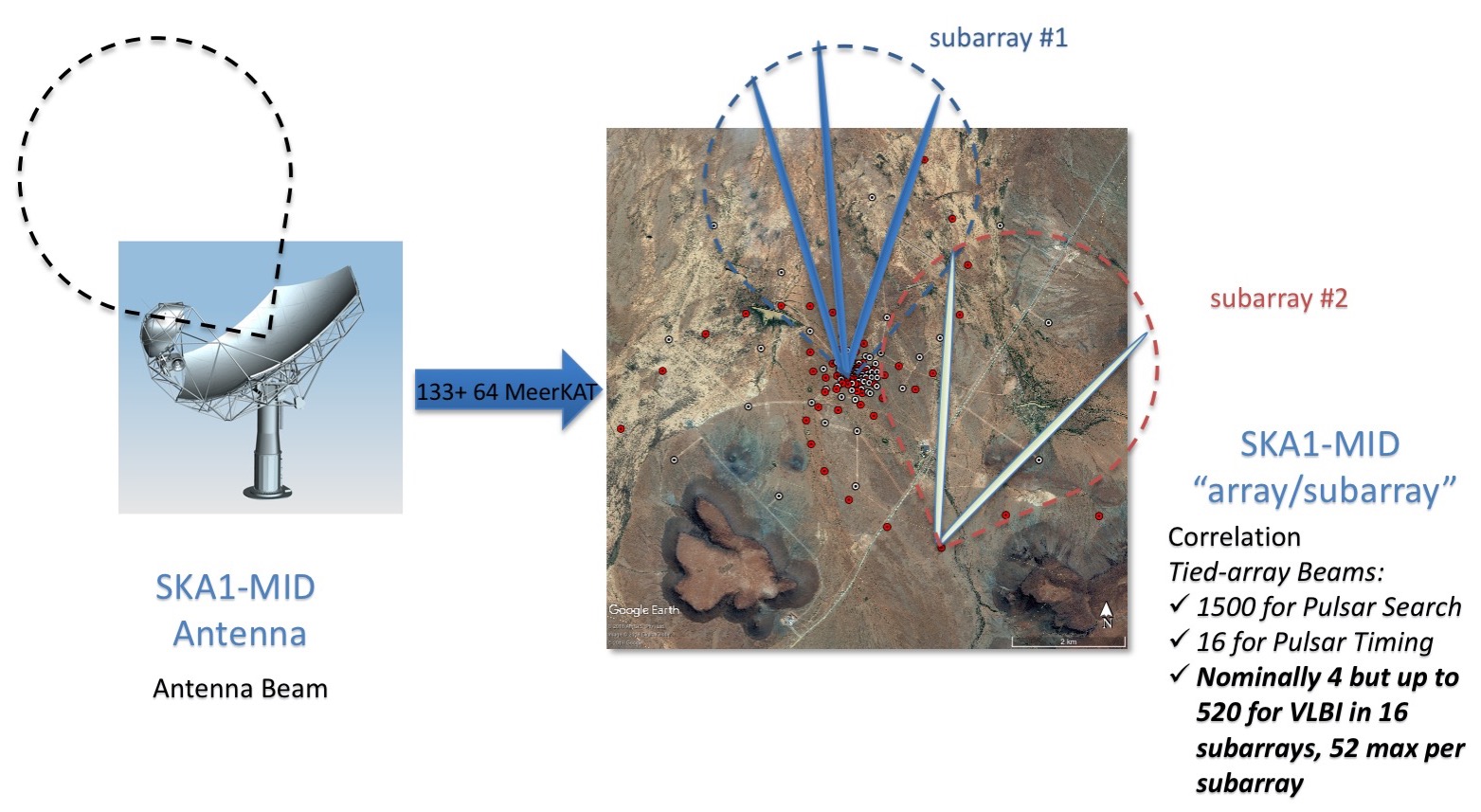} \caption{VLBI
capability for SKA1-MID telescope. The MID correlator can process up to 16 subarrays simultaneously.
The subarrays are formed by a subset of antennas, containing up to the full array (133 SKA1-MID
antennas and 64 MeerKAT antennas) or just one antenna. Up to 52 VLBI tied-array beams of 200 MHz
bandwidth per polarisation can be formed for each subarray independently, with a total of 520 beams
formed from a maximum of 16 subarrays. For larger bandwidths, fewer beams are formed per subarray,
up to a minimum of 4 VLBI beams with 2.5 GHz bandwidth per polarisation. The VLBI beams can be
pointed at any direction of the sky within the primary beam of the largest antennas used in the
subarray.} \label{fig4} \end{figure}

The VLBI beams can be formed using different subarray sizes, up to the full telescope array, with
the caveat that longer baselines will suffer from coherence losses and the beams will provide
limited FoV. The VLBI capability of MID surpasses that in the System Level 1 requirements
\cite{SKA-Req}, by providing a total of 520 VLBI beams of 200 MHz effective bandwidth in dual
polarisation from a total of 16 subarrays. Each subarray can produce a maximum of 52 beams of 200
MHz effective bandwidth, or a lesser number of beams but with larger bandwidths, e.g. 4 beams of 2.5
GHz bandwidth in dual polarisation (Fig.~\ref{fig4}).

At the 200 MHz level the RFI contamination is flagged or excised if present and polarisation leakage
is corrected. Each 200 MHz frequency slice is channelised in up to 4 dual polarisation beam channels
(i.e. subbands in VLBI jargon), with a tunable centre frequency, and bandwidths ranging from 1 to
128 MHz and up to the full 200 MHz bandwidth. Digitisation uses 2 to 16 bits per sample with Nyquist
sampling. Beam channels are formatted in VDIF packets using real representation, that are streamed
to the VLBI terminal. Power levels in the beam channels and beamforming weights are provided as part
of the metadata. Together with VLBI beams, the MID correlator provides visibility data at reduced
spectral resolution compared to normal imaging visibilities, to provide calibration solutions to
establish beam coherence and for imaging products in support of VLBI calibration.

\begin{figure}[!ht] \vspace{1.0 cm} \includegraphics[angle=0,width=6.2in]{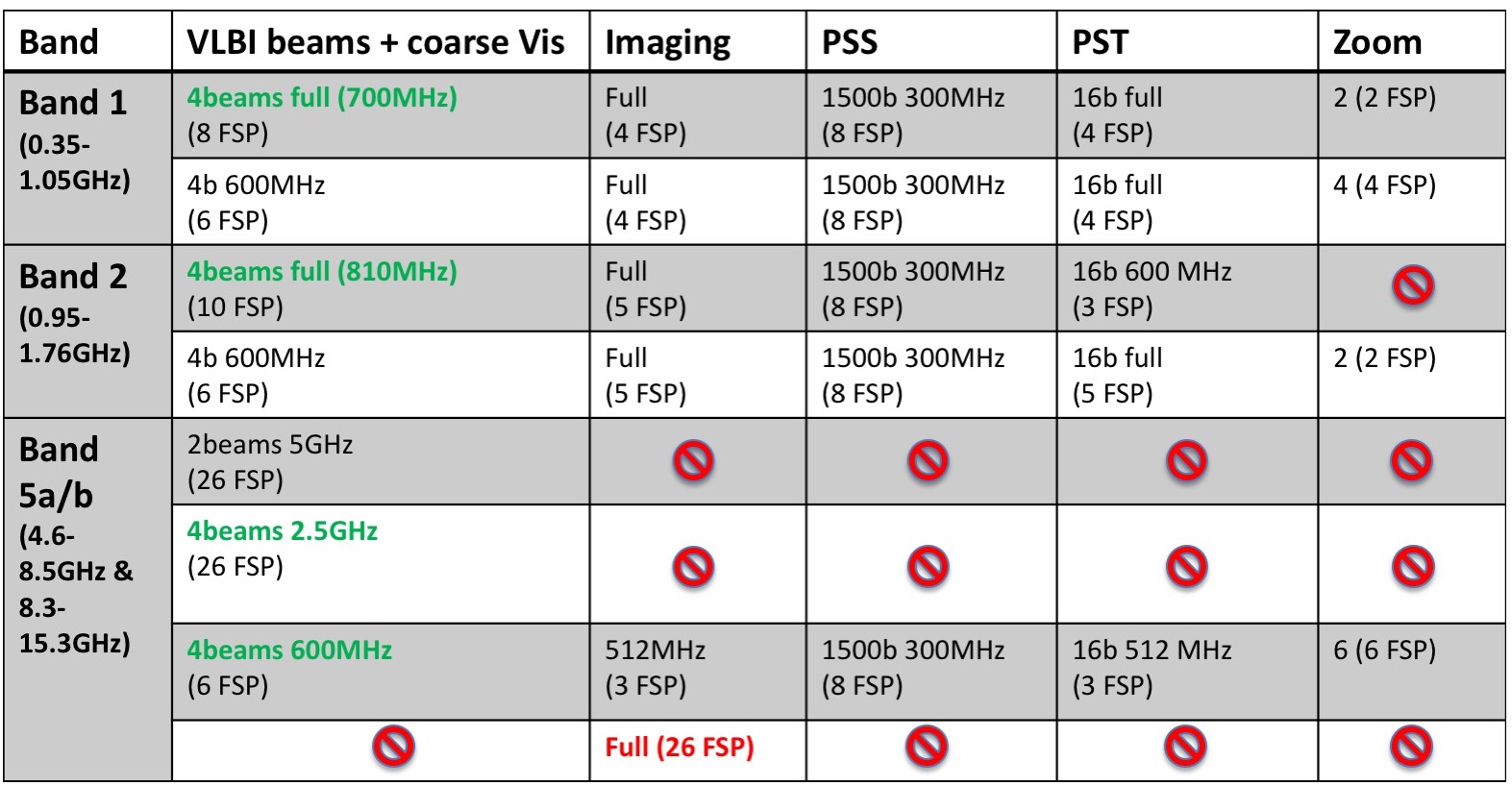}
\caption{SKA1-MID telescope observing commensality. The MID correlator provides enough processing
resources (Frequency Slice Processors, FSPs) to be able to process full Band 5 bandwidth (5 GHz
instantaneous bandwidth) for one processing mode at a time: Imaging, Pulsar Search (PSS), Pulsar
Timing (PST), VLBI, and Zoom windows for spectroscopy. For example for full bandwidth Imaging in
Band 5, all FSPs are used, not allowing for any other simultaneous processing mode. In Bands 1 and
2, where the total instantaneous bandwidth does not exceed 1 GHz, all modes can be observed
simultaneously at full bandwidth. The table shows examples of allowed commensal observations in the
different bands and the resources used for each mode, i.e. number of FSPs.} \label{fig5}\end{figure}

The MID correlator initial deployment will provide enough processing resources to be able to process
the full 5 GHz instantaneous Band 5 bandwidth for just one processing mode at a time. For example
for full bandwidth imaging in Band 5, all 26 available FSPs are required. Full simultaneity
(commensality) of the different observing modes is achieved in all bands for each subarray for
moderate observing bandwidths (Fig.~\ref{fig5}).

\subsection{VLBI with the SKA1-LOW}

The SKA1-LOW correlator and beamformer design also presents an FX-type FPGA based correlator. The
correlator is not limited by processing resources as the bandwidth to be processed is much narrower
(300 MHz bandwidth between 50-350 MHz) \cite{SKA-Base}. The resources required are available to
provide all different processing modes, Imaging and tied-array beamforming, for each subarray
simultaneously, for up to 16 independent subarrays.

\begin{figure}[!ht] \vspace{1.0 cm} \includegraphics[angle=0,width=6.2in]{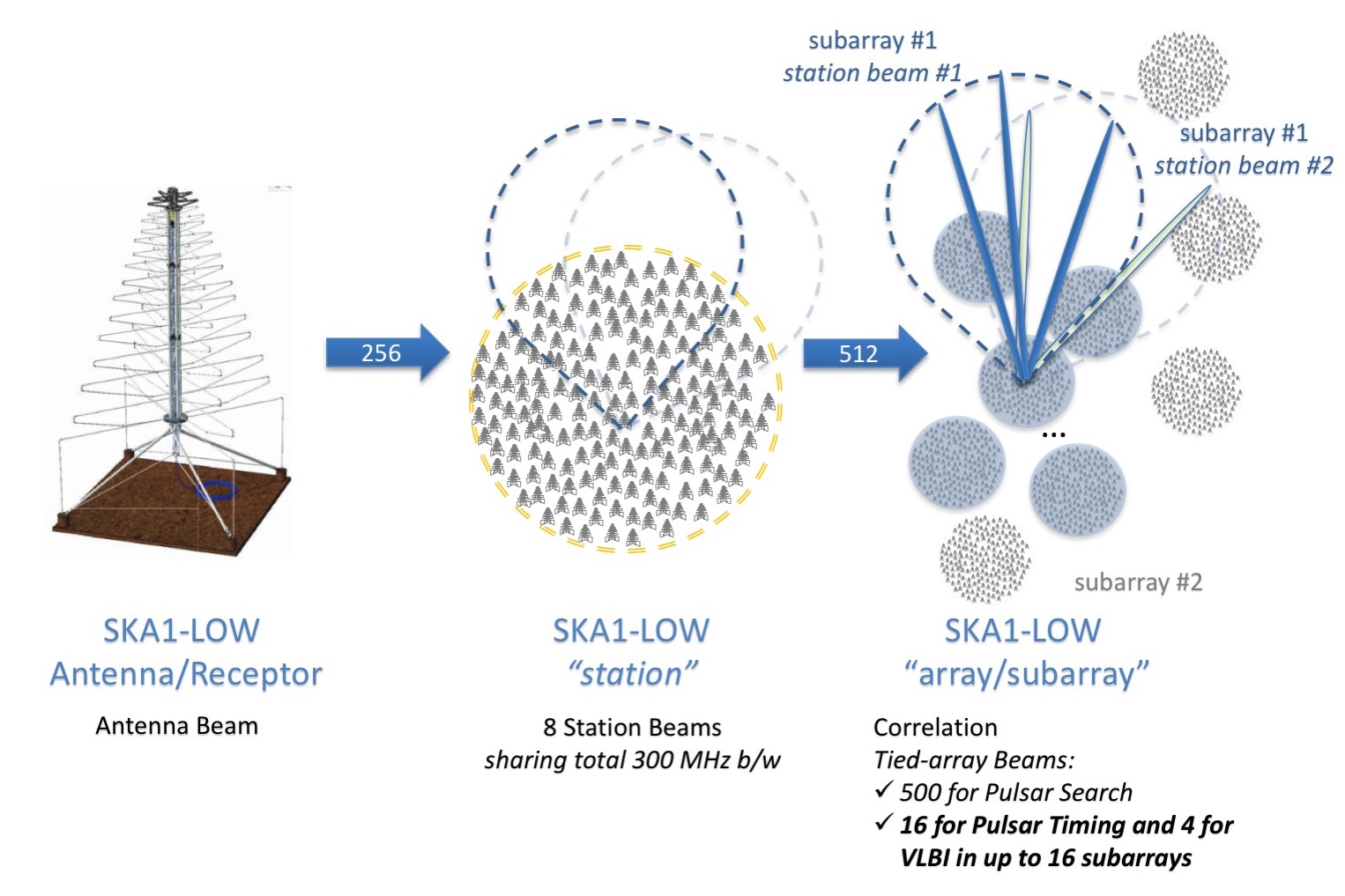} \caption{VLBI
capability for SKA1-LOW telescope. The LOW correlator can process up to 16 subarrays simultaneously.
The subarrays are formed by a subset of stations, containing up to the full array (512 stations) or
just one station. Each station is made of 256 log-periodic antennas. The total instantaneous
processed bandwidth is 300 MHz divided amongst up to 8 different stations beams pointed in different
directions. Up to 4 VLBI tied-array beams of 256 MHz bandwidth per polarisation can be formed in
total from one or up to 4 subarrays using just one station beam. If a VLBI beam needs to be pointed
outside the station beam, a different station beam could be used to generate it by trading off some
of the available 300 MHz bandwidth. In the event that Pulsar Timing is not being observed
commensally with VLBI, up to 16 VLBI beams of 256 MHz bandwidth per polarisation could be formed
from one or up to 16 subarrays.} \label{fig6} \end{figure}

The LOW correlator is required to form, in independent directions, 500 beams for the Pulsar Search
(PSS), 16 beams for Pulsar Timing (PST) and 4 beams for VLBI. The beams can be produced from the
same subarray or distributed amongst different subarrays (Fig.~\ref{fig6}). The maximum subarray
size for tied-array beamforming is up to 20 km in diameter. The beamforming process for PST and VLBI
beams is exactly the same, sharing the resources for up to 16 beams. In principle VLBI requires the
use of 4 PST beams but it could process more if they are not used by PST.

The VLBI beams have a maximum bandwidth of 256 MHz per polarisation. Before beamforming, at the LFAA
fine channel level, RFI contamination is flagged or excised and the polarisation leakage is
corrected. Each VLBI beam is channelised in up to 4 dual polarisation beam channels contiguous in
frequency, with a tunable centre frequency and bandwidths ranging from 1 to 64 MHz. Digitisation
uses 2 to 8 bits per sample and up to 2x Nyquist oversampling. Beam channels are formatted in VDIF
packets using real representation, that are streamed to the VLBI terminal. Power levels in the beam
channels and beamforming weights are provided as part of the metadata.

\subsection{VLBI element of the SKA1}

The VLBI element falls under the responsibility of an international VLBI Consortium. It will be
composed of a VLBI terminal with VDIF recorders, monitor and control software, and the necessary
scripts to carry out the observations. To allow easy access for the VLBI Consortium to the VLBI
terminal, and to not impose additional burden to the very restricted RFI control at the telescopes
sites, the VLBI terminal will be installed at the Science Processing Centres (SPC), located in Perth
and Cape Town for SKA1-LOW and SKA1-MID, respectively.

\begin{figure}[!ht] \vspace{1.0 cm} \includegraphics[angle=0,width=6.2in]{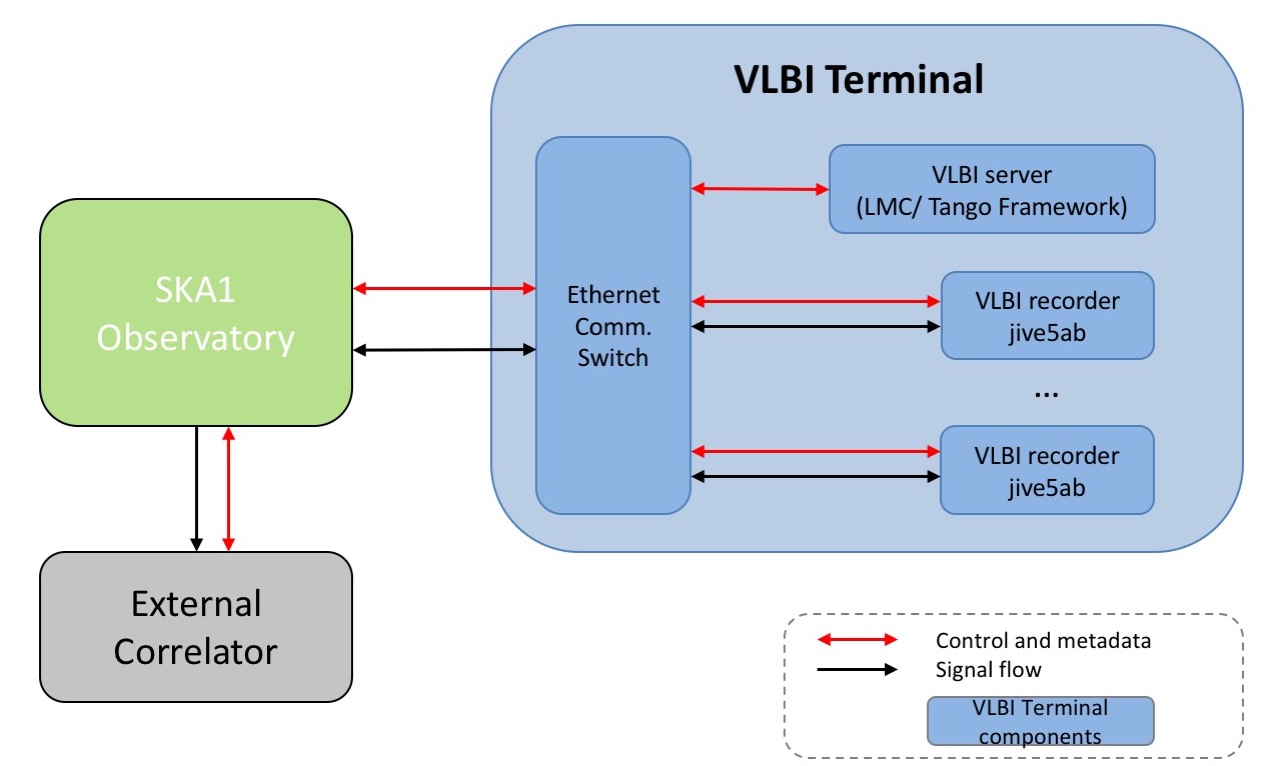}
\caption{VLBI Terminal description.} \label{fig7} \end{figure}

The VLBI terminal will be comprised of a Commercial Off-The-Shelf (COTS) VLBI server, a COTS 100GE
Ethernet Switch and one or several COTS VLBI recorders (Fig.~\ref{fig7}). The exact number of
recorders depends on the recorder performance and on the required recording data rate, with the aim
of supporting 400 Gbps (MID) and 100 Gbps (LOW) maximum data rates from the SKA1 telescopes with the
agreed interfaces. The data rate could easily surpass this planned capability, mainly for MID Band
5, therefore the VLBI equipment and interfaces will be upgraded during the lifetime of the SKA
Observatory.

The Ethernet switch will provide bi-directional communication with the SKA1 Observatory and from
there, with the outside world. The ethernet switch will receive the VLBI VDIF packets from the SKA
correlators, and either send them to the External correlator in real-time for e-VLBI observing or
stream them to the VLBI recorders for subsequent playback at a convenient time. It will also
communicate with the Telescope Manager for monitor and control tasks and subscription to metadata.
The communication with the outside world also provides access for developers for maintenance tasks
and upgrades.

The VLBI VDIF recorders selected for this design are FlexBuff type recorders but other types of
compatible VDIF recorders may be used in conjunction, or as an alternative (e.g. Mark 6 Haystack
recorders or NAOJ OCTAVE-families). FlexBuff recorders provide a flexible and standard COTS solution
for simultaneous receive, buffer (record) and transmission of VLBI data streams. Requirements for a
FlexBuff vary depending on the buffer capacity and the sustained data rate that needs to be achieved
\cite{FlexBuff}. The VLBI recorders will be controlled using the jive5ab open source control
software \cite{jive5ab}.

The VLBI server will implement SKA1 Local Monitoring and Control (LMC) within the Tango open source
framework adopted by the SKA1 Observatory. The LMC will be responsible for Monitor and Control of
the VLBI terminal and for subscription to the appropriate metadata, as well as logging events and
sending alarms to the Telescope Manager. The VLBI server will implement an additional bespoke
application to extract the metadata and generate the observing experiment log required by the
External correlator that will be recorded in the FlexBuff recorders along with the data. The LMC
will also implement a translator for jive5ab commands to control the recorders and use standard
Tango translators for the Linux server and the Ethernet switch.

\section{SKA-VLBI Science Use Cases}

The SKA1 Scientific Use Cases document \cite{SKA-Use} contains six science use cases that request the SKA-VLBI
capability (out of a total of 58 cases). These were formulated before the detailed VLBI
implementation was known. It is time to reconsider and update some of these use cases, as well as
other, new use cases that could exploit the unique SKA-VLBI capabilities. Notably, there are no use
cases for the SKA1-LOW telescope, because VLBI was not considered for this telescope in the early
days of the project. This is because of the limited number of other telescopes available in the
50-350 MHz frequency range. JUMPING JIVE is engaging with the SKA Science Working Groups, Focus
groups, and members of the VLBI community for the preparation of additional VLBI science cases that
properly reflect the recent evolution of high-profile VLBI science with a view to exploit the full
capability of the SKA1 telescopes. 

The \lq\lq SKA General Science meeting and Key Science Workshop'' (April 8-12, 2019) and the \lq\lq
SKA-VLBI Key Science Projects and Operations Workshop'' (October 14-17, 2019)
will be ideal forums for discussions towards the inclusion of VLBI science in the future SKA Key
Science Projects \cite{SKA-VLBI18}.

\acknowledgments{The \lq\lq VLBI with the SKA'' work package is part of the JUMPING JIVE project, 
that has received funding from the European Union's Horizon 2020 research and innovation programme 
under grant agreement No 730884.}


\begin{thebibliography}{99}

\bibitem{SKA-VLBI15} Z. Paragi, L. Godfrey, C. Reynolds et al., {\it Very Long Baseline
Interferometry with the SKA}, Proceedings of Science, PoS[AASKA14]143, 2015

\bibitem{SKA-VLBI18} Z. Paragi, A. Chrysostomou, C. Garc\'ia-Mir\'o, {\it SKA-VLBI Key Science
Programmes}, these proceedings

\bibitem{SKA-Ops} R.C. Bolton, A. Chrysostomou, G.R. Davis et al., {\it SKA1 Operational Concept
Document}, SKA-TEL-SKO-0000307, Rev 3, 2018

\bibitem{SKA-Base} P. Dewdney, W. Turner, R. Braun et al., {\it SKA1 System Baseline Design},
SKA-TEL-SKO-0000002, Rev 3, 2016

\bibitem{SKA-Req} M. Caiazzo, {\it SKA Phase 1 System Requirements Specification},
SKA-TEL-SKO-0000008, Rev 11, 2017

\bibitem{FlexBuff} E.Turtiainen, M. Uunila, A. Mujunen, J. Ritakari, {\it Hardware design document
for simultaneous I/O storage elements}, NEXPReS Deliverable D8.2,
http://www.jive.nl/nexpres/lib/exe/fetch.php?media=nexpres:2011-02-28\_wp8-d8.2.pdf, 2011

\bibitem{jive5ab} H. Verkouter, {\it Jive5ab command set 1.10},
http://www.jive.nl/~verkout/evlbi/jive5ab-documentation-1.10.pdf, 2018

\bibitem{SKA-Use} J. Wagg, {\it SKA1 Scientific Use Cases}, SKA-TEL-SKO-0000015, Rev 3, 2016


%
%
\end{thebibliography}
\end{document}